# ESO's Exposure Time Calculator 2.0

Henri M.J. Boffin*, Jakob Vinther, Lars K. Lundin, Gurvan Bazin
*ESO, Karl-Schwarzschild-str. 2, 85748 Garching, Germany*

## ABSTRACT

Users of the La Silla Paranal Observatory have to rely on the ESO Exposure Time Calculator (ETC) to prepare their observations. A project has been started at ESO to modernise the ETC, based on a python backend and an *angular*-based front-end. The ETC 2.0 will have a programmatic interface to enable the results to be included in an automated quality control loop and to communicate with the Phase 1 proposal preparation and the Phase 2 observation preparation tools, the ESO science archive, as well as with scripts runs by external users or instruments. The first version of an ETC 2.0 has been released for the 4MOST instrument and further versions will be released over the next years for all new La Silla, VLT and ELT instruments. The ETC of the current La Silla and VLT instruments will also be migrated progressively, with improved instrument description.

## 1. ETC?

Telescope time on large facilities, such as those operated by ESO at the La Silla Paranal Observatory, is expensive! It is thus important to optimise their use for the desired science case. As most instruments do not have real-time measurements of the signal achieved while exposing, it is essential to *predict* the signal-to-noise ratio (SNR) for a given exposure time. This is the role of the **Exposure Time Calculator (ETC)**. There exists one ETC for each main mode of all ESO La Silla Paranal instruments (Fig. 1; *https://www.eso.org/observing/etc/*).

### 1.1 Current situation

Users of the La Silla Paranal Observatory (LPO) who submit an observing proposal, through the Phase 1 (P1; proposal preparation) interface, need to provide an estimate of the required execution time of the proposed observations. Similarly, when preparing their observations, during Phase 2 (P2), they need to indicate in the Observation Blocks the exact requested observing time to reach the SNR, or the limiting magnitude[1], they have determined to be necessary to reach their science goals. In both cases, this is done through the ETC, which is a web-based application that provides the user with an estimate of the required science exposure time needed to achieve the given SNR for a given instrument at a given wavelength with the given instrument configuration and observing conditions. The current ETC (except for 4MOST, see below) provides an *HTML/Javascript*-based interface, tailored to the chosen instrument, and consist of two pages (Fig. 2). The observation parameters page presents the entry fields and widgets for the target information, expected atmospheric conditions, instrument configuration, observation parameters such as exposure time or signal-to-noise, and output selection. An "Apply" button submits the parameters to the model executed on an ESO server. The result page (Fig. 3) presents the computed results, including number of counts for the object and the sky, signal-to-noise ratios, instrument efficiencies, point spread function (PSF) size, etc. The optional graphs are displayed with *Javascript* plots, allowing interactive manipulation. The results are also provided in ASCII, PDF and GIF formats for further analysis and printing. Finally, a summary of the input parameters is appended to the result page.

* hboffin@eso.org

[1] Or precision on the visibilities and closure phases for the interferometric instruments, or standard deviation of Stokes parameters for polarimetry. When SNR is mentioned in this paper, it should be understood in its wider sense, thus including limiting magnitude and precision on interferometric observables or Stokes parameters.

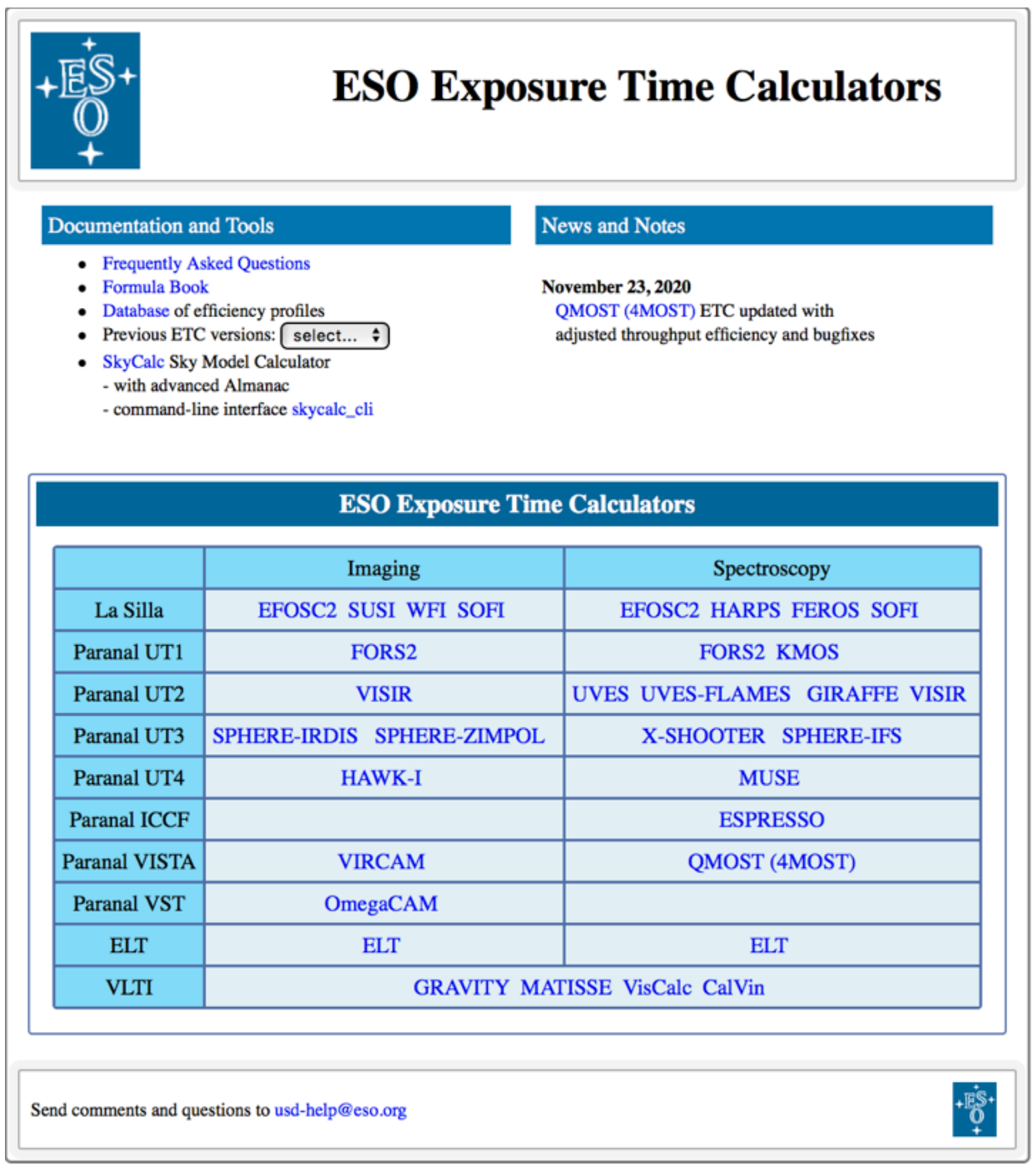

**Figure 1.** The web page of all ESO La Silla Paranal ETCs.

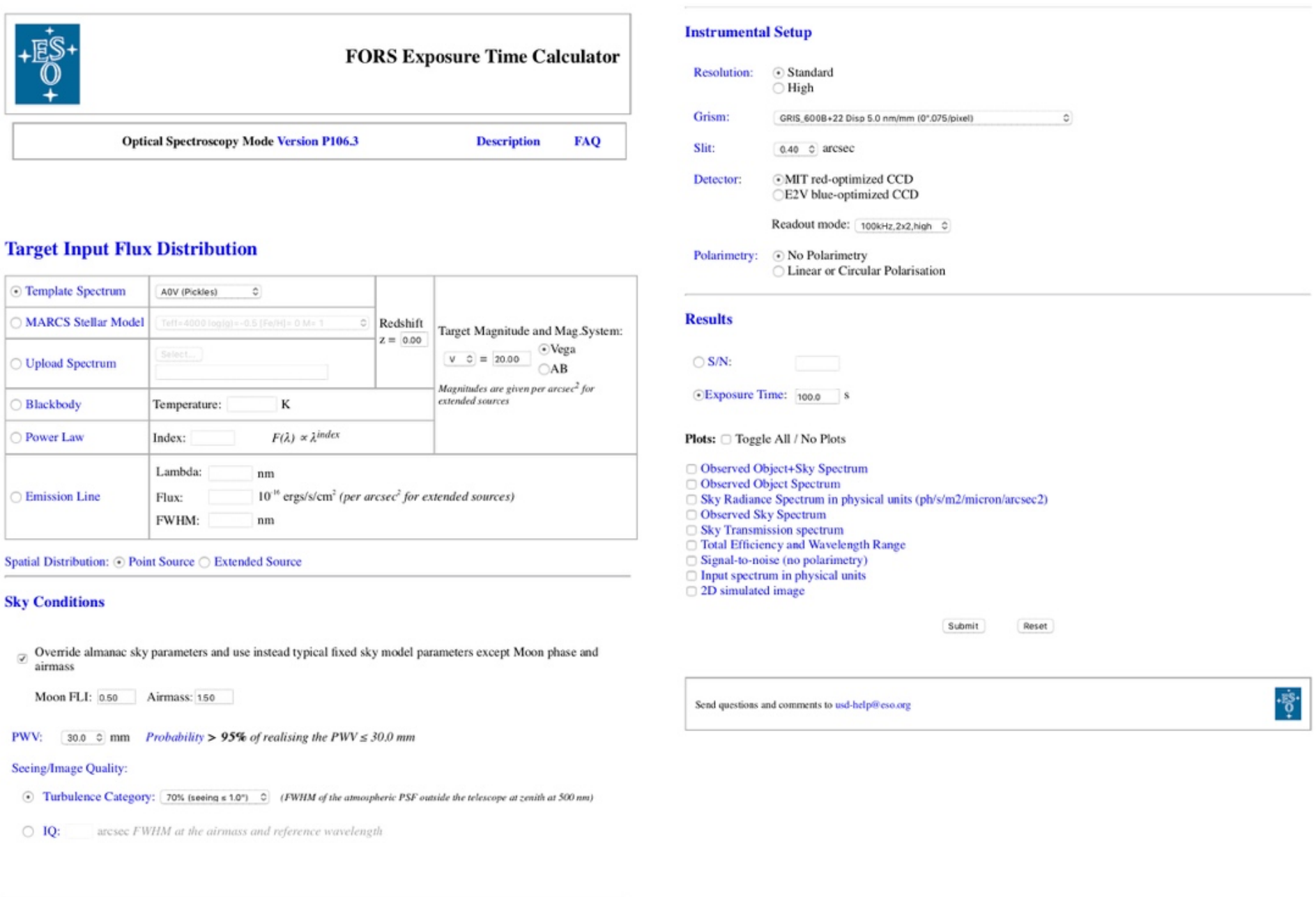

**Figure 2.** Screenshot of the ETC web page, showing part of the input required to compute the SNR.

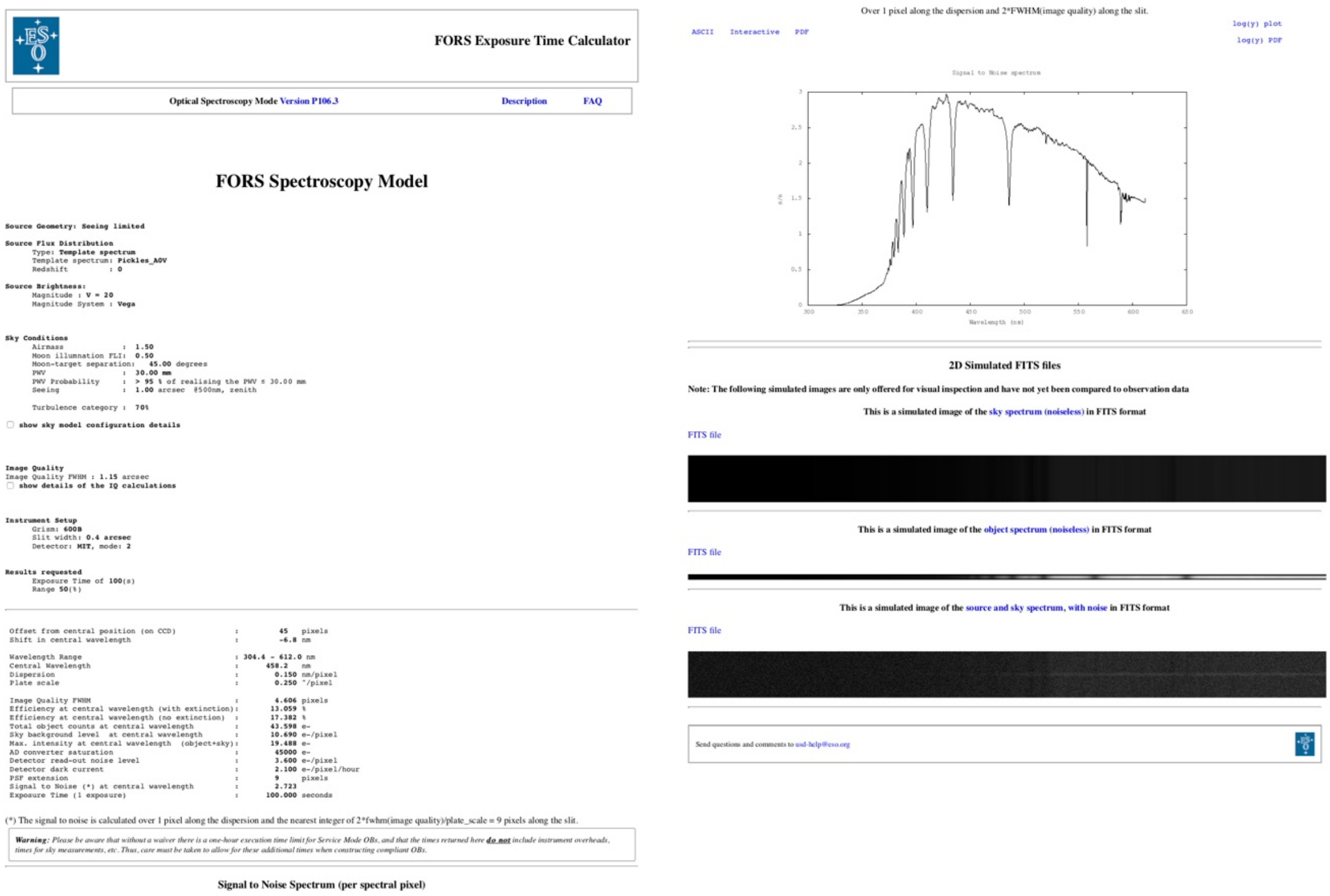

**Figure 3.** The output of the current, version 1.0, ETC for the spectroscopic mode of FORS2.

The ETC is a crucial part of the Very Large Telescope (VLT) dataflow (and, later, Extremely Large Telescope – ELT) as it enables to obtain reliable exposure times and, when accurate enough, allows one to optimise the time used on sky. The ETC must reflect the versatility of observing modes offered for each instrument. Physical models of the atmosphere, telescope and instrument systems, as well as the astronomical sources are implemented through appropriate parameterisations in the ETC. This requires regular updates, following changes in the instruments or upgrades, as well as for new instruments. The ETCs are also an essential part of the quality control (QC) loop as they can compare certain parameters from actual observations with the predictions from the ETC.

In the current system, there is no interface between P1, P2 and the ETC, nor any formal feedback with QC. All calculations in the ETC are done by the user externally to P1 and P2, who has to duplicate (triplicate!) the work. The user is then providing in the proposal a description of how they estimated the total requested time. This explanation has to be decoded and verified by ESO staff to validate the proposal and the observation blocks. Likewise, there is currently no closed loop with the master calibrations as produced and certified by the QC group. They reflect current measurements of instrument performance which however is not updated in the ETC. ETC predictions of instrument performance, on the other hand, are not compared to the instrument performance plots on the Health Check monitor.

## 1.2 Specific Issues

- The current ETC system was designed and implemented in the late nineties, that is, it is based on old technology. The requirements at the time did not include integration with other tools.
- The current code base does not systematically include input parameter validation nor runtime exception handling. The current code base does not include unit tests.
- The scope is limited to consider only SNR (or equivalently the exposure time), not addressing overhead times or specific observation date/time.

- The photometric throughput is modelled with wavelength dependent efficiency curves measured or estimated, usually adjusted to obtain the observed performance by applying "fudge factor" profiles. The ETC throughput models are adjusted occasionally, but not systematically. There is no systematic coupling with Quality Control nor the data archive, and no connection to the Phase1 or Phase2 systems.
- The current ETC system is becoming difficult to maintain and is no more compatible with the current P1 and P2 tools. Currently, the ETC is a server/client CGI web application (separate input <==> output web pages), with a C/C++ backend and a front-end based on static *HTML4* templates. *JavaScript* has later been added to the front-end, as more dynamic interactive features in input as well as output pages have been requested. Several additional technologies are involved on the server side, including shell-script, *PHP*, *python*, *Gnuplot* and a home-grown server-side macro language to manipulate *HTML*.
- Error (variability/uncertainty) estimates are not included in the current ETC model.
- It is not possible for an ETC user to save/restore/share an ETC browser session.
- Previous versions of ETCs are made available on the web by maintaining virtual machines with clones of the full historical web server including the ETC installation, executional environment and dependen-cies, for each of the historical ETC web server versions.
- The current system only supports web browser clients and does not offer a web programmatic interface (API). It is only possible to execute the ETC with one target per execution. Running the ETC for a set of configurations or a series of object characteristics implies running it many times and copying the results elsewhere by hand.
- A realistic PSF model is crucial for the ETC accuracy, in particular the adaptive optics (AO) modes. In the current ETC, the AO-PSF models for different AO-facilities and AO-instruments are very diverging in terms of model, implementation and accuracy. For non-AO modes, a Gaussian PSF model is always used. The relevance of this may be questioned for some instrument, such as MUSE, for which the PSF shape is far from Gaussian.
- The current ETCs do not have a library of line-spread functions (LSF). Therefore, the SNR in spectrographic ETC is given per spectral pixel (or spectral bin) as opposed to SNR per spectral resolution element, which is what the users are usually interested in; currently the ETC users must do their own manual estimation to get the integrated SNR.

## 2. MOVING FORWARD

Given the above, it was decided to embark on the ETC 2.0 project, aiming at modernising the LPO ETC. Recently, the Phase 2 observation preparation tool (formerly P2PP, now P2) has been migrated from a standalone Java application to a web interface in order to enhance the user friendliness of the tool, ease its maintenance – in particular its upgrades and new releases, and make it more robust. The phase 1 tool (P1) also underwent a complete re-engineering and is entirely web based. It is foreseen that the ETC will use the exact same instrument models as P1 and P2 (described by the Instrument Packages, IPs).

As stated above, currently, users need to specify many details about their observations during the Phase 1 proposal preparation, and then repeat the exercise (using different tools) when preparing the observations for the approved programme (e.g., specification of the instrument setup and observing strategy), in Phase 2. In both cases, they will need to use the ETC and have to recopy the various values and parameters, thereby creating frustration and increasing the risk of errors. ESO staff need to verify these parameters and thus, once more, repeat the whole exercise, leading to loss of efficiency and wasting of time and resources. Ideally, the user should not have to enter more than once the information about what and how they want to observe, and should be able to retrieve from the ETC the required total execution time to achieve their science, taking into account the overheads.

It is therefore desirable to have a tool, the ETC, that is fully integrated into P1 and P2, to provide the users with a unique tool and improve dramatically the workflow, as well as to reduce the risk of errors and the need for checking the Phase 2 material. It should be possible to run ETC simulations by reading the parameters from P1 and P2 into ETC. After editing the ETC input parameters a button should enable to update the P1 or P2 fields and the OB parameters to reflect those used in the ETC for the selected list of OBs.

The proposed project will have the following *scientific* benefits:
a) Users will be able to use information provided by other databases on the Internet and won't need to repeat providing information entered for the Phase 1 proposal preparation, and they therefore can use their time to define carefully the most important science related parameters. They will experience an integrated approach between Phase 1 and Phase 2 and will be able to reuse information for following periods.
b) The ETCs will have up-to-date configuration, using the IPs, ensuring users will be sure to derive values only for available instrument configurations.
c) The ETCs will have improved instrument models, thereby ensuring a better adequacy between estimated and real values, optimising the use of the telescopes and ensuring the users get the needed signal for their science.
d) The ETCs will have the versions of the instruments at a given time, allowing going back in time to compare values between estimated and real values (i.e., IP versioning).

The proposed project will have the following *operational* benefits:
a) Improving the overall integration of the many different tools for observation preparation is expected to reduce the maintenance and user support loads.
b) Enabling more automatic checks of the Phase 1 content based on the ETC, therefore reducing some manual work for operations staff, in particular during the technical feasibilities of the proposals done by Paranal.
c) Remove the need to duplicate information about the instruments, as the IPs will be used and avoid having inconsistent instrument setup between the ETC and the IP.
d) Simplify the interactions with the users who need to prepare their proposals and observations.
e) Allow quality check of the instruments by comparing values between estimated and real values, in real time.

## 3. SOFTWARE ARCHITECTURE PRINCIPLES

The new ETC solution is developed in terms of web applications backed by a server-side application doing the numerical calculations of the ETC model. The server-side application has access to data in a relational database. The following architecture principles apply:

- All required functionality is developed in terms of web applications in order to benefit from continuous "zero install" deployment, allowing to silently and quickly push out bug fixes, improvements and new features to the users. No desktop application is developed.
- Server-side business logic is separated from its user interface representation into a stand-alone backend, which is accessed via a published RESTful application programming interface and which can be comprehensively tested and scaled for required performance. Although the primary application on the client side is a web browser handled by an ETC front end this architecture will support web clients of any kind – including command-line or programmatic interfaces – to call the ETC backend through the published API.
- The web user interfaces shall be exclusively implemented on the client-side using Google's Angular framework (*angular.io*), thereby de-centralising the GUI presentation computation power into the users' browsers.

The backend is based on a *Django* web server that links to the ETC computation engine itself, to the Advanced Cerro Paranal Sky Model (*eso.org/sci/software/pipelines/skytools/skymodel*) and the Almanac calculations, and to the required *MySQL* databases. As stated above, the front-end is based on the Angular framework.

## 4. IMPLEMENTATION

The first, stand-alone version, of the ETC 2.0 for the 4MOST instrument to be installed on the VISTA telescope in Paranal has already been deployed and is available at *etc.eso.org/qmost* (Fig. 4). The very similar *look & feel* to the ESO P1 and P2 tools is apparent. By comparison with Fig. 2 and 3, it can clearly be seen that the look is modernised and made more compact, showing both the input and the resulting plots side by side.

The data interface between the ETC 2.0 front-end and backend has the data interchange format JSON. Thus, the input parameters assigned in the front-end is streamed in that format via an HTTPS POST request to the backend. Similarly, the backend returns the calculated results in JSON format to the front-end, which can extract and display various information as requested and present spectral, spatial or temporal vector data in form of dynamic plots using the interactive *JavaScript* library *HighCharts*. Both the input and output JSON data can be downloaded as files, are human-readable and can be manually or programmatically modified by advanced users.

It will thus be possible to request and retrieve the data from the backend to use and produce plots in the user's preferred software, but most importantly, as indicated above, this means that it is possible to query the ETC engine in a scripting way, using the appropriate application programming interface, for example in a python script or a command-line tool such as *curl* or *wget*. The future interaction with P1 and P2 will be using a similar data interchange scheme.

Please note that the described direct interface with the backend is presently not yet documented and not yet fully implemented, for example it is missing parameter validation and error handling. Later we will provide utilities to ease the scripting interface with the backend, similar to the already available *skycalc_cli* (*https://www.eso.org/observing/etc/doc/skycalc/helpskycalccli.html*) command-line interface to the skycalc sky model.

As Fig. 4 demonstrates, the front-end allows to define the input parameters in a series of tabs, covering the target, the sky conditions, the required seeing or image quality, the requested exposure time or signal-to-noise ratio, and to select the plots to be shown.

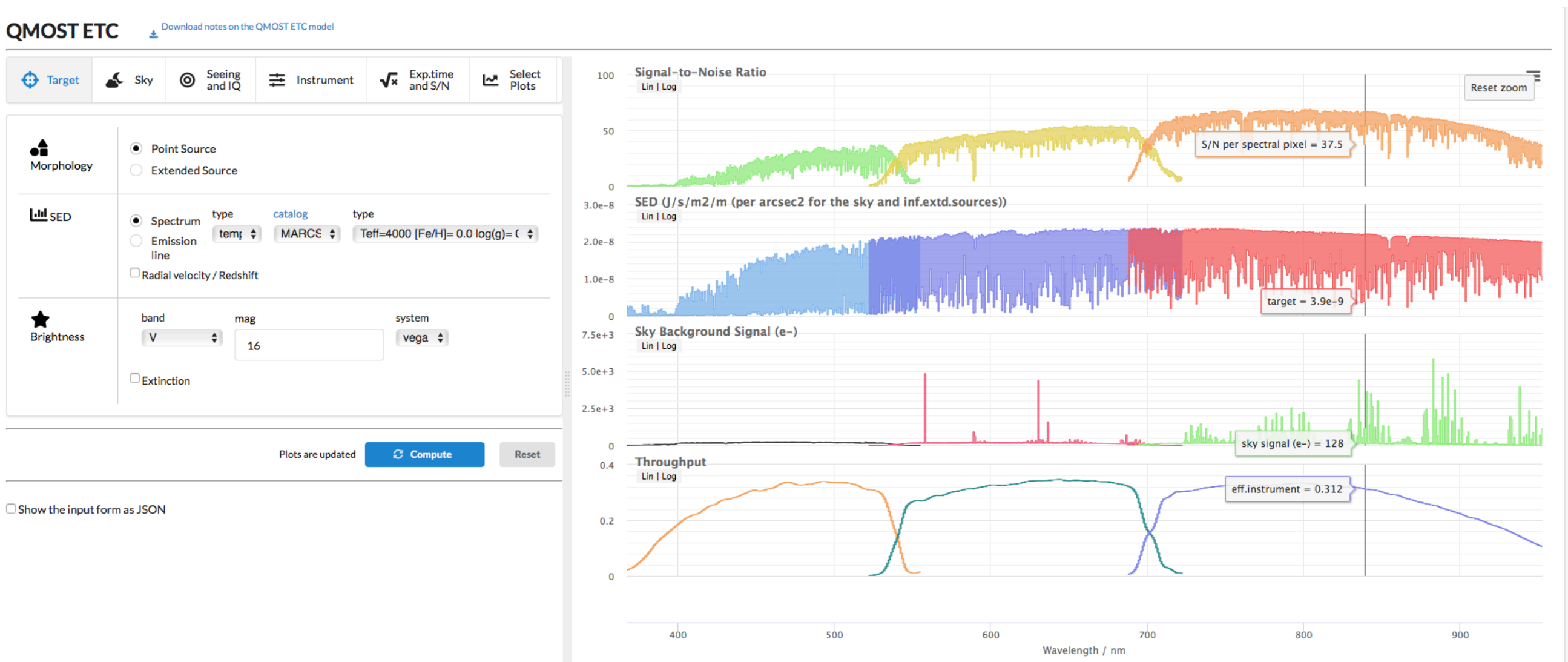

**Figure 4.** The first ETC 2.0, for the 4MOST multi-object spectrograph, has already been deployed.

### 4.1 Target definition

As illustrated in Fig. 5, which shows the ETC 2.0 of the CRIRES+ prototype, a target is defined by its

- *Designation*: this is ideally a name belonging to the dictionary of nomenclature of celestial objects so that a Name Resolver can be used (e.g., from CDS Simbad, which is currently implemented) to automatically retrieve useful information about it. This is not compulsory, however (and sometimes not possible), and any designation can be used.
- *Coordinates*: Right Ascension and Declination in a given, recognised, system. This can be automatically filled in by the Name Resolver and is used to compute the barycentric correction (see below) and the airmass and Moon-target distance, if one so wishes.
- *Morphology*: point source or extended source. In the latter case, the 2D profile should be specified and some analytic forms can be used, such as infinite, Sersic profile, power law, etc.
- *Spectral energy distribution* (SED), i.e., the flux of the source, $F_\lambda$, in erg/s/A/cm$^2$ or W/m$^2$/μm as a function of wavelength, $\lambda$, or $F_\nu$ as a function of the frequency, $\nu$. This can be a user input file, an analytic spectrum, an emission line, or a spectrum from one of the available databases (e.g., MARCS, Kurucz and Pickles for stars, Kinney for galaxies, etc.).
- *Brightness*: Target magnitude in a given system (Vega or AB) or flux in Jansky. Magnitudes and flux are given per arcsec$^2$ for extended sources.
- *Redshift or radial velocity*, and if needed, an additional barycentric correction.
- *Extinction* in a given magnitude band. The typical Milky Way $R$=3.1 curve is then used to adjust the extinction at other wavelengths.
- A *background*, which is characterised by a morphology, a SED and a brightness, as defined above.

We welcome ideas for additional functional forms for the morphology or for the template spectra. For the latter, however, it is important that they are provided in a wide wavelength range to cover all the LPO instruments, as well as with enough spectral resolution to cover all possible science cases. We are also looking at implementing in a robust fashion that the other parameters of the target, such as morphology, SED, brightness, redshift or radial velocity are automatically retrieved by the Name Resolver.

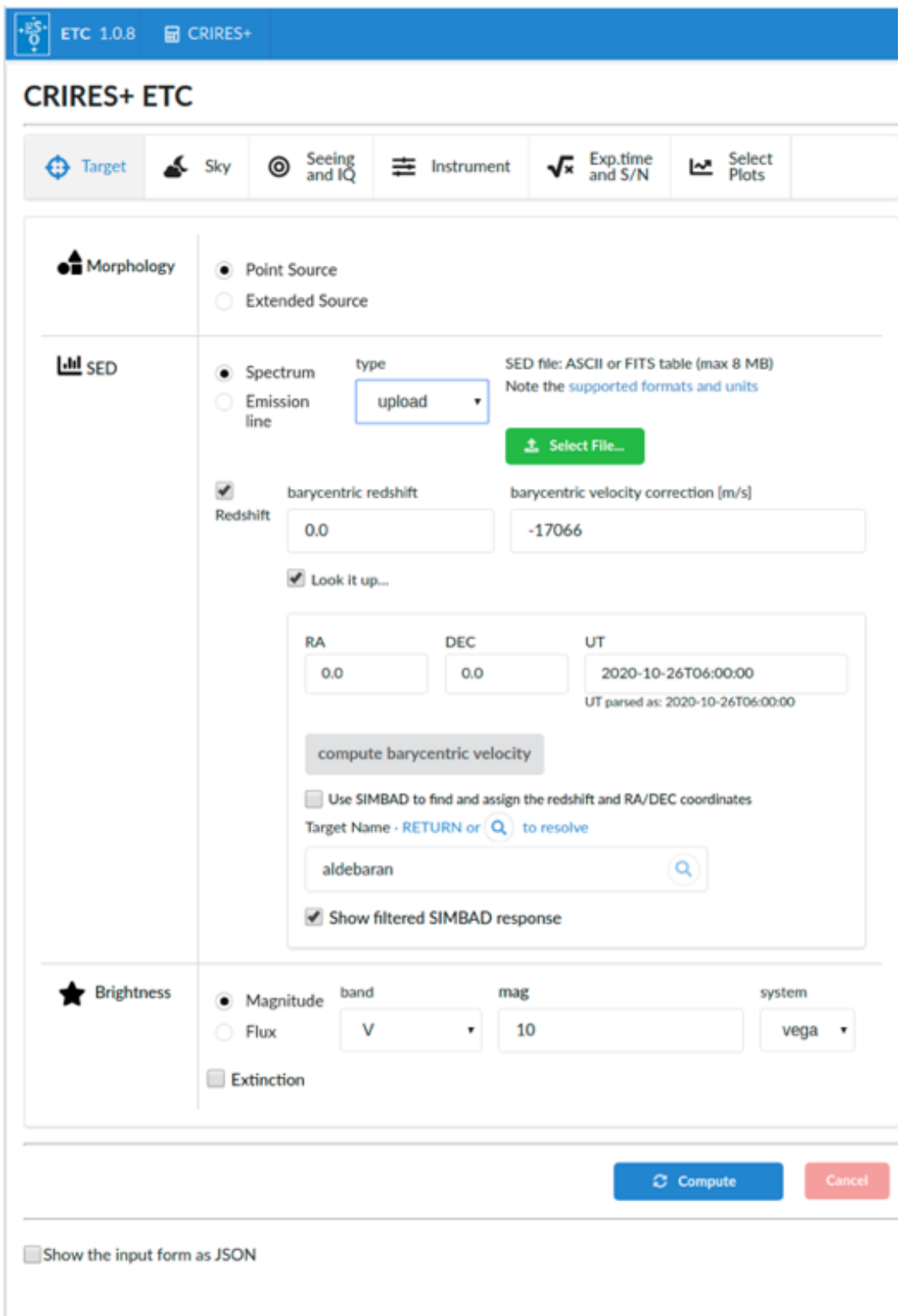

**Figure 5.** Some fields of the target tab expanded in the CRIRES+ ETC 2.0.

### 4.2 Sky

The Sky tab (Fig. 6) allows the users to provide the information about the ambient conditions relevant for the observations. In the current implementation, they are still minimal, and the users can indicate the airmass of the observations, the Moon phase (FLI), the Moon-target angular distance, as well as the precipitable water vapour (mm). In the future, an almanac will be included, in the same way as is done in the current ETC 1.0, that can compute the three first parameters directly from the coordinates of the target (retrieved, if possible, from a Name Resolver) and the expected date of observations.

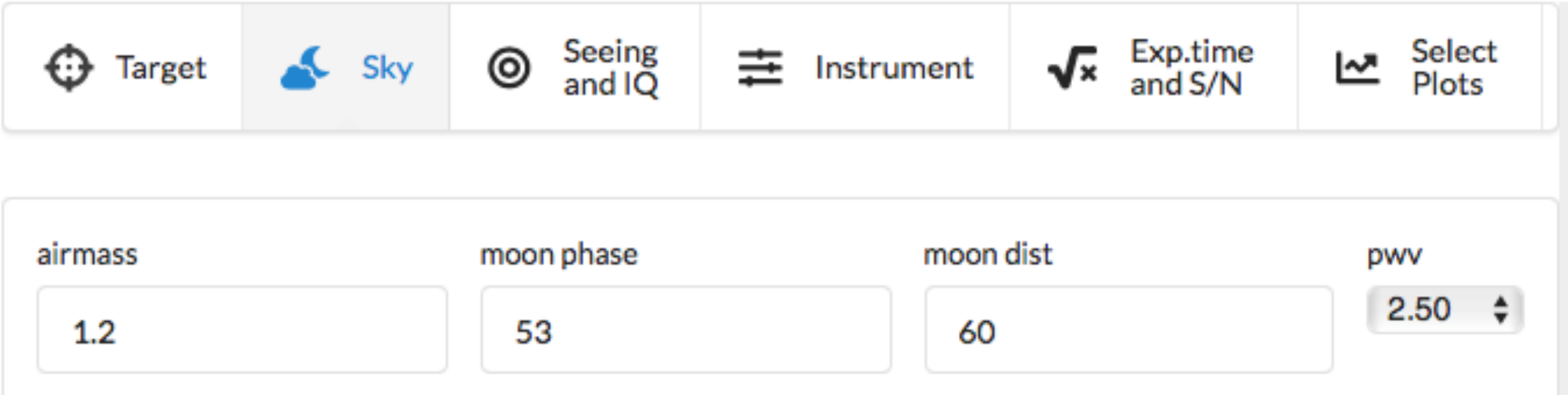

**Figure 6.** The Sky tab expanded in the CRIRES+ ETC 2.0.

### 4.3 Seeing and IQ

The additional ambient parameter that is required for the ETC to be able to compute the SNR as a function of exposure time is the seeing (measured in the *V* band at zenith), or more generally, the turbulence category (Fig. 7). ESO has indeed recently introduced a new constraint, the atmospheric turbulence constraint. This constraint generalises the classical seeing constraint used so far and takes into account the coherence time or the fraction of turbulence in the ground layer, for instruments that need it, in addition to the classical seeing. These additions are important constraints for instruments using adaptive optics and need to be taken into account when scheduling observations to ensure that the science goals can be achieved. The turbulence category is defined by its probability of occurrence, from 10%, meaning the top 10% of the best conditions in La Silla or Paranal (the most demanding category), to 100%, meaning any conditions.

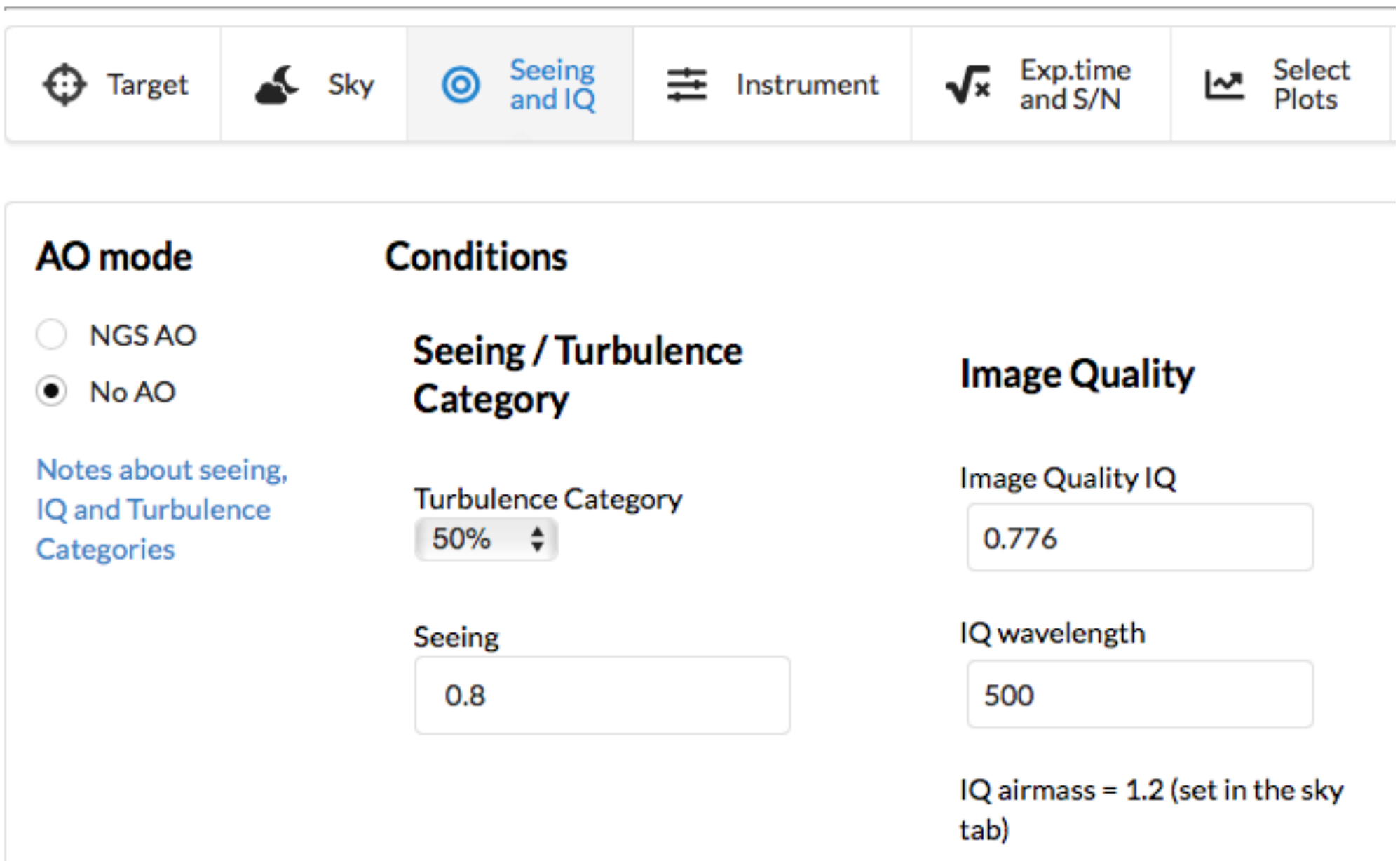

**Figure 7.** The Seeing and IQ tab expanded in the CRIRES+ ETC 2.0.

At a given seeing, one can also associate an Image Quality (IQ), defined as the full width at half maximum (FWHM) of long-exposure stellar images – it is a property of the images obtained in the focal plane of an instrument mounted on a telescope observing through the atmosphere. It is therefore a quantity measured at the (requested) airmass and wavelength of observation and that is also computed on the fly by the ETC.

## 4.4 Instrument

The next tab to be filled in is the *Instrument* one (Fig. 8). This is where the user will specify the modes to be used for the observations. This can be very simple, as in 4MOST, where there is only the choice between the low (LRS) and high (HRS) spectral resolution modes, or more complex, when one needs to choose a specific setting, an order, a slit width or whether polarimetry will be done or not.

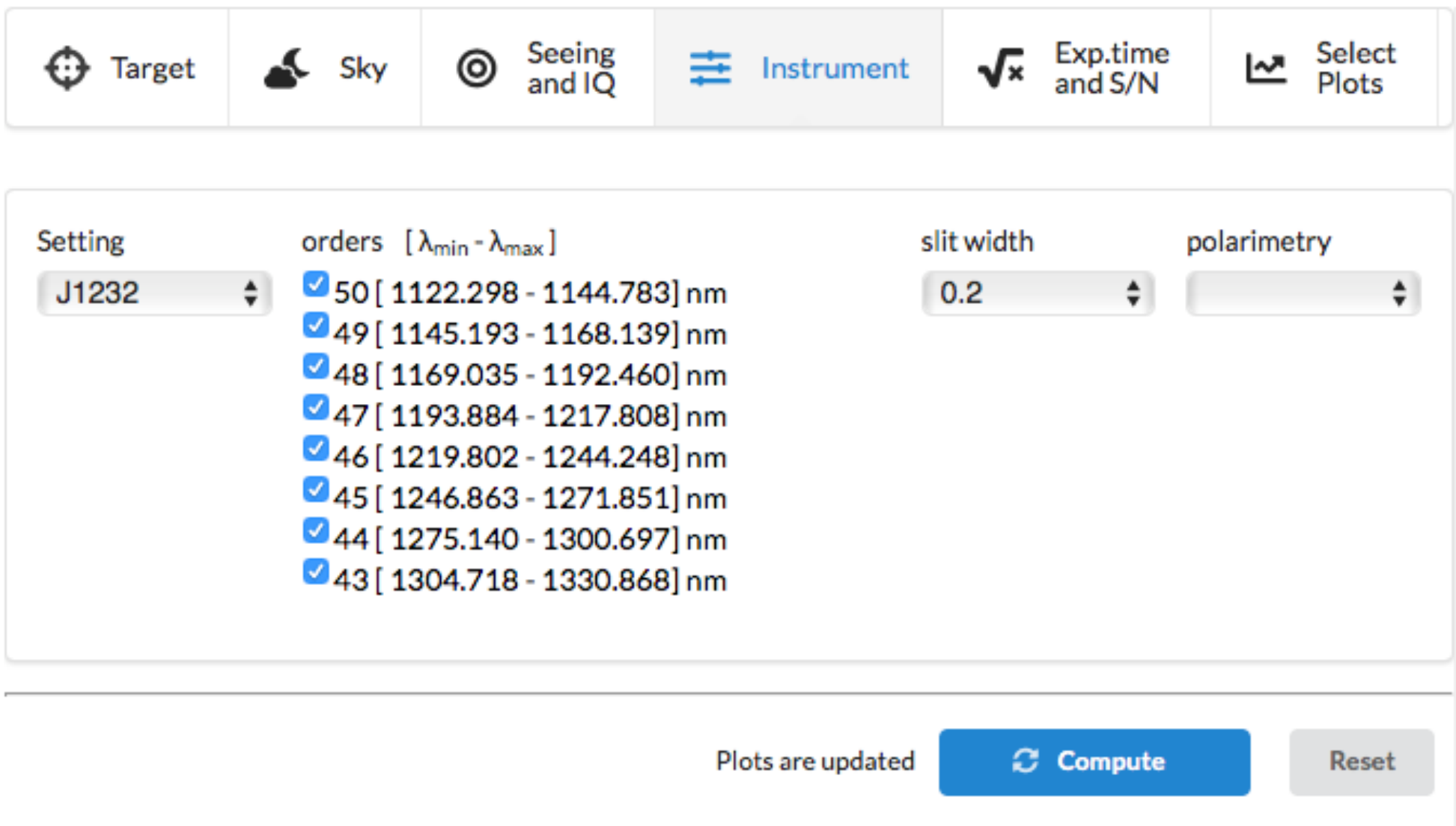

**Figure 8.** The instrument tab expanded in the CRIRES+ ETC 2.0.

### 4.5 Exposure time and SNR

Finally, the users will have to define either the desired exposure time (that can be split in a number of sub-integrations) or the requested signal-to-noise ratio (SNR) – see Fig. 9.

### 4.6 Plots

With all the parameters introduced, the users can then finally choose which plots they want to see (Fig. 10). The plots are adjustable, and one can zoom in a given spectral range.

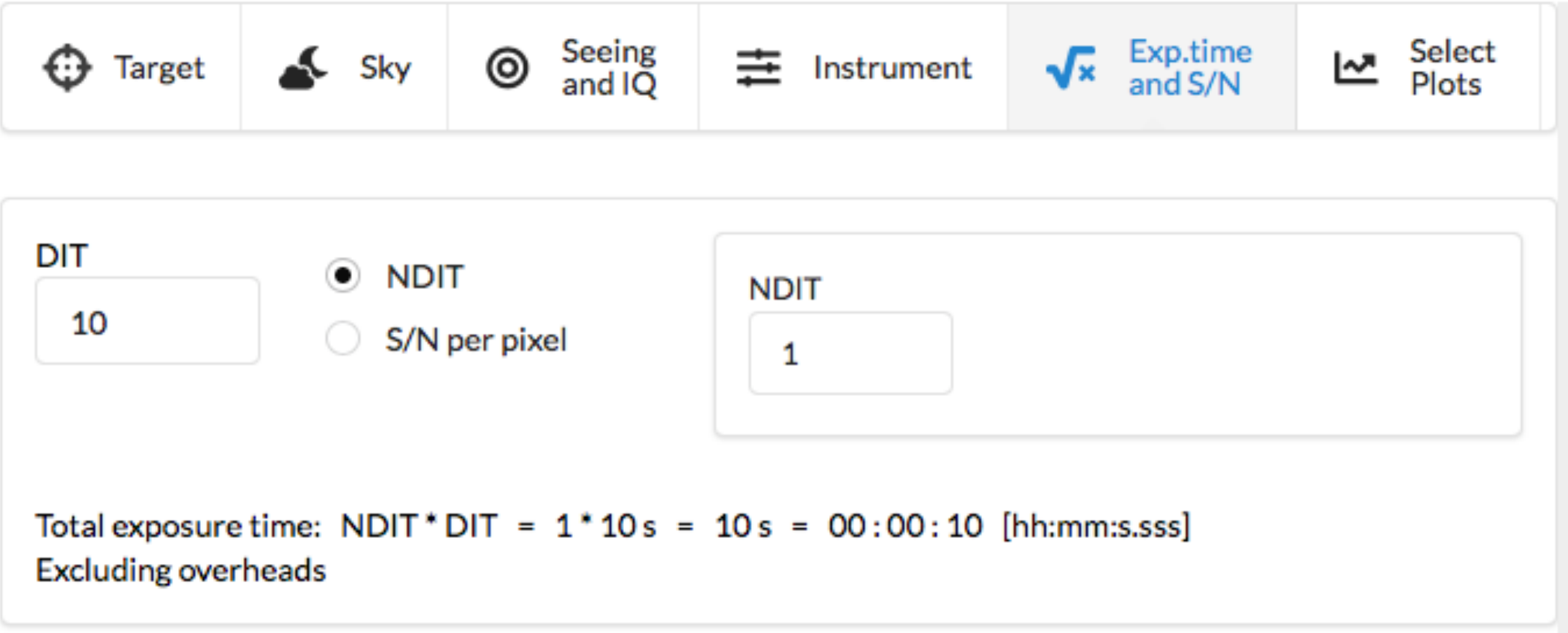

**Figure 9.** The exposure time and SNR tab expanded in the CRIRES+ ETC 2.0.

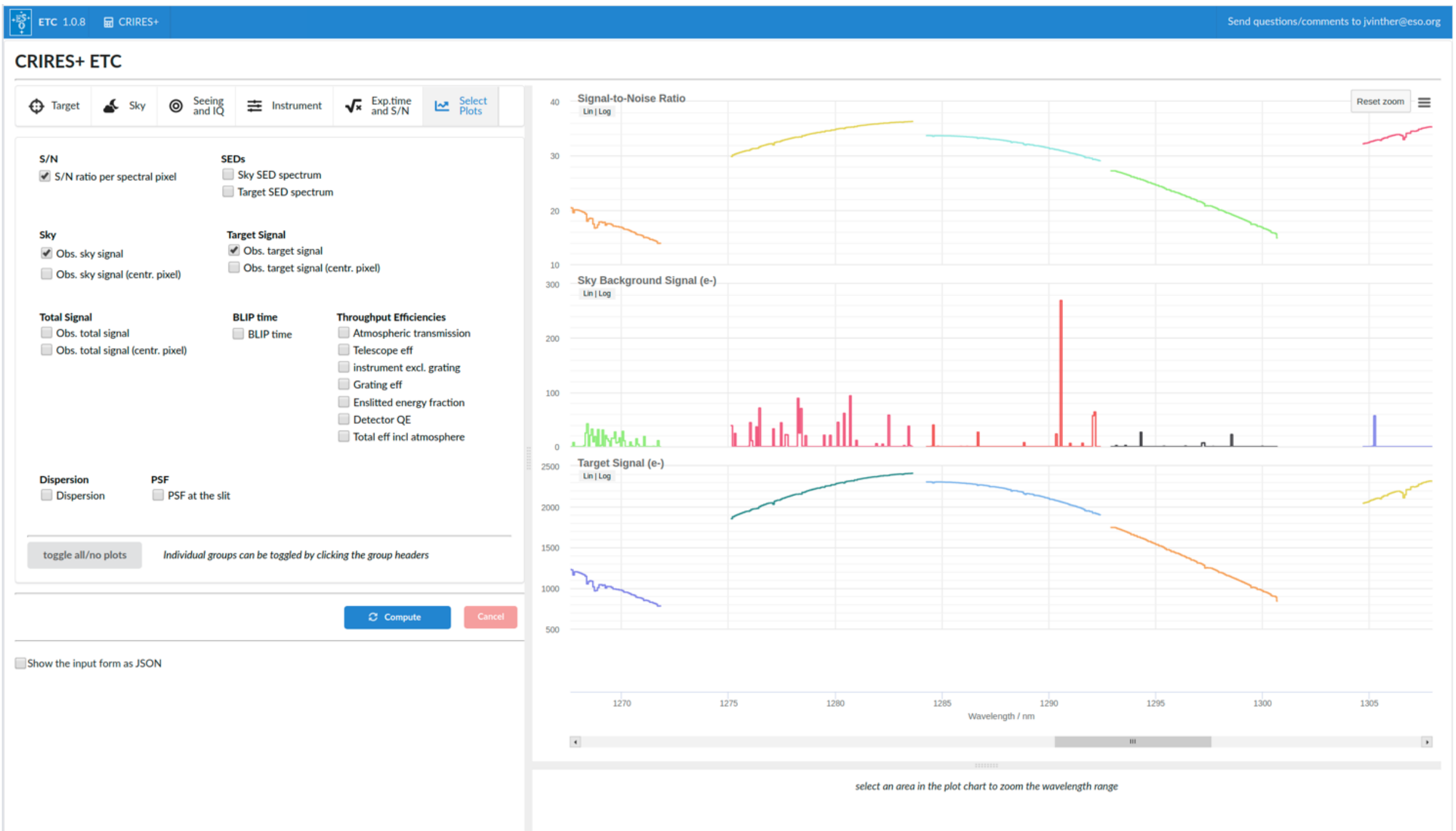

**Figure 10.** The *Select Plots* tab for the CRIRES+ ETC 2.0.

## 5. FINAL THOUGHTS

In the framework of the ETC 2.0 project, it is foreseen that the Exposure Time Calculators (ETC) of all current and future La Silla Paranal instruments will be moved to the ETC 2.0 framework, allowing a complete integration with P1 and P2. This implies porting the current models of these instruments to the python back end engine. The new ETCs should also be as accurate as possible and allow for a "real time" update of the model components, so as to always reflect the state of the instrument, using a feedback loop with Quality Control at the Observatory and relying on data in the ESO Science Archive whenever possible. To this aim, all the current ETC models and their components will be reviewed and updated if need be. Work has already started on porting the FORS2 ETC to the new framework and it is foreseen that all LPO instruments will be ported within the next three years. In addition to the 4MOST ETC 2.0 which is already public, the ETC 2.0 for the soon-to-come at Paranal CRIRES+ instrument is already well developed. A preview is available on *etctestpub.eso.org/observing/etc/crires*. Please note that at this stage it is an early preview; the performance predictions have not been validated with measurements obtained in commissioning and nothing can be guaranteed.